\documentclass[letterpaper, 11pt]{article}
\usepackage{amsmath}
\usepackage{amsfonts}

\begin{document}

%\begin{center}
%{\Large {\bf Many Hard Examples in Exact Phase Transitions\\[0.3cm]
%{\normalsize with Application to Generating Hard Satisfiable Instances\footnote{%
%{\small This research was partially supported by the National Key Basic Research
%Program (973 Program) of China under Grant No. G1999032701.}}}}}\\[0.5cm]
%{\large Ke Xu and Wei Li}

\begin{center}
{\Large {\bf Many Hard Examples in Exact Phase Transitions with\\[0.3cm]
Application to Generating Hard Satisfiable Instances\footnote{%
{\small This research was partially supported by the National Key Basic Research
Program (973 Program) of China under Grant No. G1999032701 and Special Funds
for Authors of National Excellent Doctoral Dissertations of China under Grant No.
200241. Preliminary version of this paper appeared as Technical Report cs.CC/0302001 
of CoRR in Feb. 2003. }}}}\\[0.5cm]
{\large Ke Xu and Wei Li}

\bigskip
%{\setlength{\parskip}{0.cm}
National Lab of Software Development Environment

Department of Computer Science

Beihang University, Beijing 100083, China

Email:\{kexu,liwei\}@nlsde.buaa.edu.cn

\end{center}

\begin{quotation}
{\noindent {\small {\bf Abstract.} This paper first analyzes the resolution
complexity of two random CSP models (i.e. Model RB/RD) for which we can
establish the existence of phase transitions and identify the threshold
points exactly. By encoding CSPs into CNF formulas, it is proved that
almost all instances of Model RB/RD have no tree-like resolution proofs of
less than exponential size. Thus, we not only introduce new families of CNF
formulas hard for resolution, which is a central task of Proof-Complexity
theory, but also propose models with both many hard instances and exact 
phase transitions. Then, the implications of such models are addressed.
It is shown both theoretically and experimentally that an application of 
Model RB/RD might be in the generation of hard satisfiable instances, which is not only 
of practical importance but also related to some open problems in cryptography 
such as generating one-way functions. Subsequently, a further theoretical support for
the generation method is shown by establishing exponential lower bounds 
on the complexity of solving random satisfiable and forced satisfiable 
instances of RB/RD near the threshold. Finally, conclusions are presented, 
as well as a detailed comparison of Model RB/RD with the Hamiltonian cycle problem and 
random 3-SAT, which, respectively, exhibit three different kinds of phase transition 
behavior in NP-complete problems.}}
\end{quotation}

\bigskip

\noindent {\large {\bf 1. Introduction }}

\smallskip

\noindent Over the past ten years, the study of phase transition
phenomena has been one of the most exciting areas in computer science and
artificial intelligence. Numerous empirical studies suggest that for many
NP-complete problems, as a parameter is varied, there is a sharp transition
from 1 to 0 at a threshold point with respect to the probability of a random
instance being soluble. More interestingly, the hardest instances to solve
are concentrated in the sharp transition region. As well known, finding ways
to generate hard instances for a problem is important both for understanding
the complexity of the problem and for providing challenging benchmarks for
experimental evaluation of algorithms [12]. So the finding of phase
transition phenomena in computer science not only gives a new method to
generate hard instances but also provides useful insights into the study of
computational complexity from a new perspective.

Although tremendous progress has been made in the study of phase
transitions, there is still some lack of research about the connections
between the threshold phenomena and the generation of hard instances,
especially from a theoretical point of view. For example, some problems can
be used to generate hard instances but the existence of phase transitions in
such problems has not been proved. One such an example is the well-studied
random 3-SAT. A theoretical result by Chv\'{a}tal and Szemer\'{e}di [10]
shows that for random 3-SAT, no short proofs exists in general, which means
that almost all proofs for this problem require exponential resolution
lengths. Experimental results further indicate that instances from the phase
transition region of random 3-SAT tend to be particularly hard to solve
[25]. Since the early 1990's, considerable efforts have been put into random
3-SAT, but until now, the existence of the phase transition phenomenon in
this problem has not been established, although recently, Friedgut [14] made
tremendous progress in proving that the width of the phase transition region
narrows as the number of variables increases. On the other hand, for some
problems with proved phase transitions, it was found either theoretically or
experimentally that instances generated by these problems are easy to solve
or easy in general. Such examples include random 2-SAT, Hamiltonian cycle
problem and random 2+$p$-SAT ($0<p\leq 0.4$). For random 2-SAT, Chv\'{a}tal
and Reed [11] and Goerdt [20] proved that the phase transition phenomenon will
occur when the ratio of clauses to variables is 1. But we know that 2-SAT is
in P class which can be solved in polynomial time, implying that random
2-SAT can not be used to generate hard instances. For the Hamiltonian cycle
problem which is NP-compete, Koml\'{o}s and Szemer\'{e}di [22] not only
proved the existence of the phase transition in this problem but also gave
the exact location of the transition point. However, both theoretical
results [9] and experimental results [32] suggest that generally, the
instances produced by this problem are not hard to solve. 
Different from the above two problems, random
2+$p$-SAT [30] was first proposed as an attempt to interpolate between the
polynomial time problem random 2-SAT with $p=0$ and the NP-complete problem
random 3-SAT with $p=1.$ It is not hard to see that random 2+$p$-SAT is in
fact NP-compelte for $p>0.$ The phase transition behavior in this problem
with $0<p\leq 0.4$ was established by Achlioptas et al. and the exact
location of the threshold point was also obtained [1]. But it was further
shown that random 2+$p$-SAT is essentially similar to random 2-SAT when $%
0<p\leq 0.4$ with the typical computational cost scaling linearly with the
number of variables [29].

As mentioned before, from a computational theory point of view, what
attracts people most in the study of phase transitions is the finding of
many hard instances in the phase transition region. Hence, starting from
this point, we can say that the problem models which can not be used to generate
random hard instances are not so interesting for study as random 3-SAT.
However, until now, for the models with many hard instances, e.g. random
3-SAT, the existence of phase transitions has not been established, not even
the exact location of the threshold points. So, from a theoretical
perspective, we still do not have sufficient evidence to support the
long-standing observation that there exists a close relation between the
generation of many hard instances and the threshold phenomena, although this
observation opened the door for, and has greatly advanced the study of phase
transitions in the last decade. From the discussion above, an interesting
question naturally arises: {\em whether there exist models with both
proved phase transitions and many hard instances and, if so, what are the
implications of such models.} 

Recently, to overcome the trivial asymptotic insolubility of the previous
random CSP models, Xu and Li [33] proposed a new CSP model, i.e. Model RB,
which is a revision to the standard Model B. It was proved that the phase
transitions from solubility to insolubility do exist for Model RB as the
number of variables approaches infinity. Moreover, the threshold points at
which the phase transitions occur are also known exactly. Based on previous
experiments and by relating the hardness of Model RB to Model B, it has
already been shown that Model RB abounds with hard instances in the phase
transition region. In this paper, we will first propose a random CSP model,
called Model RD, along the same line as for Model RB. Then, by encoding CSPs
into CNF formulas, we will prove that almost all instances of Model RB/RD
have no tree-like resolution proofs of less than exponential size. This
means that Model RB/RD are hard for all popular CSP algorithms because 
such algorithms are
essentially based on tree-like resolutions [24]. Therefore, we not only
introduce new families of CNF formulas hard for resolution, which is a
central task of Proof-Complexity theory, but also propose models
with both many hard instances and exact phase transitions. 
More importantly, it will be shown that an application of RB/RD 
might be in the generation of hard satisfiable instances, which is not only 
of significance for experimental studies, but also of interest to the theoretical 
computer science community.   
Finally, exponential lower bounds will be established for random satisfiable
and forced satisfiable instances of RB/RD near the threshold.

\bigskip

\noindent {\large {\bf 2. Model RB and Model RD}}

\smallskip

\noindent A {\it Constraint Satisfaction Problem}, or CSP for short, 
consists of a set of variables, a set of possible values for 
each variable (its domain) and a set
of constraints defining the allowed tuples of values for the variables
(a well-studied special case of it is SAT). 
The CSP is a fundamental problem in Artificial Intelligence, with a distinguished
history and many applications, such as in knowledge representation, scheduling
and pattern recognition. To compare the efficiency of different CSP algorithms,
some standard random CSP models have been widely used experimentally to
generate benchmark instances in the past decade. For the most widely used CSP
model (i.e. standard Model B), Achlioptas et al. [2] proved that except for a small
range of values of the constraint tightness, almost all instances generated
are unsatisfiable as the number of variables approaches infinity. This result,
as shown in [19], implies that most previous experimental results about random
CSPs are asymptotically uninteresting. However, it should be noted that
Achlioptas et al.'s result holds under the condition of fixed domain size and
so is applicable only when the number of variables is overwhelmingly larger
than the domain size. But in fact, it can be observed that the domain size,
compared to the number of variables, is not very small in most experimental
CSP studies. This, in turn, explains why there is a big gap between Achlioptas
et al.'s theoretical result and the experimental findings about the phase
transition behavior in random CSPs. Motivated by the observation above, and
to overcome the trivial asymptotic insolubility of the previous random
CSP models, Xu and Li [33] proposed an alternative CSP model as follows.

{\bf Model RB: }First, we select with repetition $m=rn\ln n$ random
constraints. Each random constraint is formed by selecting without
repetition $k$ of $n$ variables, where $k\geq 2$ is an integer. Next, for
each constraint we uniformly select without repetition $q=p\cdot d^{k}$
incompatible tuples of values, i.e., each constraint contains exactly $%
(1-p)\cdot d^{k}$ allowed tuples of values, where $d=n^{\alpha }$ is the
domain size of each variable and $\alpha >0$ is a constant.{\em \ }

Note that the way of generating random instances for Model RB is almost the
same as that for Model B. However, like the N-queens problem and Latin square,
the domain size of Model RB is not fixed but polynomial in the number of
variables. It is proved that Model RB not only avoids the trivial asymptotic
behavior but also has exact phase transitions. More precisely, the
following theorems hold for Model RB, where $\Pr (Sat)$ denotes the probability 
that a random CSP instance generated by Model RB is satisfiable.

%Xu and Li [23] further proved that the probability that a random CSP
%instance generated by Model RB is satisfiable, denoted by $\Pr (Sat),$
%exhibits phase transitions at a threshold point known exactly, i.e. the
%following theorems hold for Model RB.

{\bf Theorem 1} \ (Xu and Li [33]) Let $r_{cr}=-\frac{\alpha }{\ln (1-p)}$.
If $\alpha >\frac{1}{k}$, $0<p<1$ are two constants and $k$, $p$ satisfy the
inequality $k\geq \frac{1}{1-p}$, then
\begin{eqnarray*}
\underset{n\rightarrow \infty }{\lim }\Pr (Sat) &=&1\text{ when }r<r_{cr}, \\
\underset{n\rightarrow \infty }{\lim }\Pr (Sat) &=&0\text{ when }r>r_{cr}.
\end{eqnarray*}

{\bf Theorem 2} \ (Xu and Li [33]) Let $p_{cr}=1-e^{-\frac{\alpha }{r}}$. If
$\alpha >\frac{1}{k}$, $r>0$ are two constants and $k$, $\alpha $ and $r$
satisfy the inequality $ke^{-\frac{\alpha }{r}}\geq 1$, then
\begin{eqnarray*}
\underset{n\rightarrow \infty }{\lim }\Pr (Sat) &=&1\text{ when }p<p_{cr}, \\
\underset{n\rightarrow \infty }{\lim }\Pr (Sat) &=&0\text{ when }p>p_{cr}.
\end{eqnarray*}

As shown in [33], many instances generated following Model B in previous
experiments can also be viewed as instances of Model RB, and more importantly,
the experimental results for these instances agree well with the theoretical
predictions for Model RB. Therefore, in this sense, we can say that Model B
can still be used experimentally to produce benchmark instances. However, to
guarantee an asymptotic phase transition behavior and to generate random hard
instances, a natural and convenient way is to vary the values of CSP parameters 
under the framework of Model RB. Note that another standard CSP\
Model, i.e. Model D, is almost the same as Model B except that for every
constraint, each tuple of values is selected to be incompatible with
probability $p.$ Similarly, we can make a revision to Model D and then get a
new Model as follows.

{\bf Model RD: }First, we select with repetition $m=rn\ln n$ random
constraints. Each random constraint is formed by selecting without
repetition $k$ of $n$ variables, where $k\geq 2$ is an integer. Next, for
each constraint, from $d^{k}$ possible tuples of values, each tuple is
selected to be incompatible with probability $p$, where $d=n^{\alpha }$ is
the domain size of each variable and $\alpha >0$ is a constant.{\em \ }

Along the same line as in the proof for Model RB [33], we can easily prove
that exact phase transitions also exist for Mode RD. More precisely, Theorem
1 and Theorem 2 hold for Model RD too. In fact, it is exactly because the
differences between Model RB and Model RD are very small that many
properties hold for both of them and the proof techniques are also almost
the same. So in this paper, we will discuss both models, denoted by Model
RB/RD.

Recently, there has been a growing theoretical interest in random CSPs,
especially with respect to their phase transition behaviors [13, 16, 17, 27, 31, 35] 
and resolution complexity [18, 26, 28].
To discuss the resolution complexity of CSPs, we first need to encode a CSP
instance into a CNF formula. In this paper we will adopt the encoding method
used in [24]. For convenience, we give the outline of this method here. For
each CSP variable $u,$ we introduce $d$ propositional variables, called {\it %
domain variables}, to represent assignments of values to $u.$ There are
three sets of clauses needed in the encoding, i.e. the {\it domain clauses}
asserting that each variable must be assigned a value from its domain, the
{\it conflict clauses} excluding assignments violating constraints and
clauses asserting that each variable is assigned at most one value from its
domain.

\bigskip

\noindent {\large {\bf 3. Resolution Lower Bounds for Model RB/RD}}

\smallskip

\noindent In this section, we will analyze the resolution complexity of
unsatisfiability proofs for Model RB/RD and get the following result.

{\bf Theorem 3 \ }Let $P$ be a random CSP instance generated following Model
RB/RD. Then, almost surely, $P$ has no tree-like resolutions of length less
than 2$^{\Omega (n)}.$

When we say that a property holds almost surely it means that this property
holds with probability tending to 1 as the number of variables approaches
infinity.

The core of the proof for Theorem 3 is to show that almost surely there exists a
clause with large width in every refutation. The width of a clause $C$,
denoted by $w(C),$ is the number of variables appearing in it. The width of
a set of clauses is the maximal width of a clause in the set. The width of
deriving a clause $C$ from the formula $F,$ denoted by $w(F\vdash C)$ is
defined as the minimum of the widths of all derivations of $C$ from $F.$ So,
the width of refutations for $F$\ can be denoted by $w(F\vdash 0).$
Ben-Sasson and Wigderson [8] gave the following theorem on size-width 
relations and proposed a general strategy for proving width
lower bounds for CNF formulas.

{\bf Theorem 4 \ }(Ben-Sasson and Wigderson [8]) Let $F$ be a CNF formula
and $S_{T}(F)$ be the minimal size of a tree-like refutation. Then we have
\[
S_{T}(F)\geq 2^{(w(F\vdash 0)-w(F))}.
\]

By extending Ben-Sasson and Wigderson's strategy, Mitchell [26] proved
exponential resolution lower bounds for some random CSPs of fixed domain 
size. In what follows, to obtain  
lower bounds on width for RB/RD, we will basically use the same strategy
as in [26], but adapt it to handle random CSPs with growing domains. 
First, we prove the following local sparse property for RB/RD.

{\bf Lemma 1 }Let $P$ be a random CSP instance generated by Model RB/RD.
There is constant $c>0$ such that almost surely every sub-problem of $P$
with size $s\leq cn$ has at most $b=\beta s\ln n$ constraints, where $\beta =%
\frac{\alpha }{6k\ln \frac{1}{1-p}}.$

{\bf Proof: }As mentioned in [27], this is a standard type of argument in
random graph theory. Similarly, we consider the number of sub-problems on $s$
variables with $b=\beta s\ln n$ constraints for $0<s\leq cn.$ There are $%
\binom{n}{s}$ possible choices for the variables and $\binom{m}{b}$ for the
constraints. Given such choices, the probability that all the $b$
constraints are in the $s$ variables is not greater than $\left( \frac{s}{n}%
\right) ^{kb}.$ So, the number of such sub-problems is at most
\begin{eqnarray*}
\binom{n}{s}\binom{m}{b}\left( \frac{s}{n}\right) ^{kb} &\leq &\left( \frac{%
en}{s}\right) ^{s}\left( \frac{em}{b}\right) ^{b}\left( \frac{s}{n}\right)
^{kb} \\
&=&\left( \frac{en}{s}\right) ^{s}\left( \frac{ern\ln n}{\beta s\ln n}%
\right) ^{\beta s\ln n}\left( \frac{s}{n}\right) ^{k\beta \ln n} \\
&=&\left[ \frac{e^{1+\beta \ln n}r^{\beta \ln n}}{\beta ^{\beta \ln n}}%
\left( \frac{s}{n}\right) ^{(k-1)\beta \ln n-1}\right] ^{s}.
\end{eqnarray*}

\noindent For sufficiently large $n,$ there exists a constant $c_{1}>0$ such
that

\[
\frac{e^{1+\beta \ln n}r^{\beta \ln n}}{\beta ^{\beta \ln n}}<n^{c_{1}}.
\]

\noindent Thus we get

\[
\binom{n}{s}\binom{m}{b}\left( \frac{s}{n}\right) ^{kb}<\left[
n^{c_{1}}\left( \frac{s}{n}\right) ^{(k-1)\beta \ln n-1}\right] ^{s}.
\]

\noindent Let $c<\frac{1}{2}\exp\left(-\frac{2+c_{1}}{(k-1)\beta}\right)$ be
a positive constant.
For $0<s\leq cn,$ it follows from the above inequality that

\[
\binom{n}{s}\binom{m}{b}\left( \frac{s}{n}\right) ^{kb}<\left( \frac{1}{n^{2}%
}\right) ^{s}\leq \frac{1}{n^{2}}.
\]

\noindent Thus the expected number of such sub-problems with $s\leq cn$ is
at most

\[
\overset{cn}{\underset{s=1}{\sum }}\binom{n}{s}\binom{m}{b}\left( \frac{s}{n}%
\right) ^{kb}<\frac{1}{n^{2}}cn=o(1).
\]

\noindent This finishes the proof. \hfill $\Box $

\smallskip

The following two definitions will be of use later.

{\bf Definition 1 \ }Consider a variable $u$ and $i$ constraints associated
with $u.$ In these $i$ constraints, all the variables except $u$ have
already been assigned values from their domains. We call this an $i$-{\it %
constraint assignment tuple}, denoted by $T_{i,u}.$

{\bf Definition 2 \ }Given a variable $u$ and an $i$-constraint assignment
tuple $T_{i,u}.$ We assign a value $v$ to $u$ from its domain$.$ So, all the
variables in the $i$ constraints of $T_{i,u}$ have been assigned values. If
at least one constraint in $T_{i,u}$ is violated by these values, then we
say that {\it the} {\it value }$v${\it \ of }$u${\it \ is flawed} {\it by} $%
T_{i,u}.$ If all the values of $u$ in its domain are flawed by $T_{i,u},$
then we say that {\it the variable }$u${\it \ is flawed by} $T_{i,u},$ and $%
T_{i,u}$ is called a {\it flawed }$i${\it -constraint assignment tuple}.

\smallskip

{\bf Lemma 2 \ }Let $P$ be a random CSP instance generated by Model RB/RD.
Almost surely, there does not exist a flawed $i$-constraint assignment tuple
$T_{i,u}$ in $P$ with $i\leq 3k\beta \ln n.$

{\bf Proof: }Now consider an $i$-constraint assignment tuple $T_{i,u}$ with $%
i\leq 3k\beta \ln n.$ It is easy to see that the probability that $T_{i,u}$
is flawed increases the number of constraints $i.$ Recall that in Model RD,
for every constraint, each tuple of values is selected to be incompatible
with probability $p.$ So, given a value $v$ of $u,$ the probability that $v$
is flawed by $T_{i,u}$ is
\[
1-(1-p)^{i}.
\]

\noindent Thus the probability that all the $d=n^{\alpha }$ values of $u$
are flawed by $T_{i,u},$ i.e. the probability of $T_{i,u}$ being flawed is
\[
\left[ 1-(1-p)^{i}\right] ^{d}.
\]

\noindent Note that $\beta =\frac{\alpha }{6k\ln \frac{1}{1-p}}.$ Thus for $%
0<i\leq 3k\beta \ln n,$ we have
\begin{eqnarray*}
\Pr (T_{i,u}\text{ is flawed})|_{i\leq 3k\beta \ln n} &\leq &\left[
1-(1-p)^{3k\beta \ln n}\right] ^{n^{\alpha }} \\
&=&[1-\frac{1}{n^{\frac{\alpha }{2}}}]^{n^{\alpha }}\approx e^{-n^{\frac{%
\alpha }{2}}}.
\end{eqnarray*}

\noindent The above analysis only applies to Model RD. For Model RB, such an
analysis is much more complicated, and so we leave it in the appendix.
Recall that there are $n$ variables and $m=rn\ln n$ constraints. So the
number of possible choices for $i$-constraint assignment tuples is at most
\[
n\binom{m}{i}d^{(k-1)i}.
\]

\noindent For $i\leq 3k\beta \ln n,$ when $n$ is sufficiently large, there
exists a constant $c_{2}>0$ such that
\begin{eqnarray*}
n\binom{m}{i}d^{(k-1)i} &=&n\binom{rn\ln n}{i}n^{(k-1)\alpha i}\leq n\binom{%
rn\ln n}{3k\beta \ln n}n^{3(k-1)\alpha k\beta \ln n} \\
&\leq &n\left( \frac{ern\ln n}{3k\beta \ln n}\right) ^{3k\beta \ln
n}n^{3(k-1)\alpha k\beta \ln n}<e^{c_{2}\ln ^{2}n}.
\end{eqnarray*}

\noindent Thus the expected number of flawed $i$-constraint assignment
tuples with $i\leq 3k\beta \ln n$ is at most
\begin{eqnarray*}
\overset{3k\beta \ln n}{\underset{i=1}{\sum }}n\binom{m}{i}d^{(k-1)i}\Pr
(T_{i,u}\text{ is flawed}) &<&e^{c_{2}\ln ^{2}n}\overset{3k\beta \ln n}{%
\underset{i=1}{\sum }}\Pr (T_{i,u}\text{ is flawed}) \\
&=&e^{c_{2}\ln ^{2}n}\cdot O(e^{-n^{\frac{\alpha }{2}}})\cdot 3k\beta \ln n
\\
&=&o(1).
\end{eqnarray*}

\noindent This implies that almost surely, there does not exist a variable $%
u $ and an $i$-constraint assignment tuple $T_{i,u}$ with $i\leq 3k\beta \ln
n$ such that $u$ is flawed by $T_{i,u}.$ This is exactly what we need and so
we are done. \hfill $\Box $

\smallskip

{\bf Lemma 3 }Let $P$ be a random CSP instance generated by Model RB/RD.
Almost surely, every sub-problem of $P$ with size at most $cn$ is
satisfiable.

{\bf Proof: }Here by the size of a problem we mean the number of variables
in this problem. We will prove this lemma by contradiction. Assume that we
have an unsatisfiable sub-problem of size at most $cn.$ Thus we can get a
minimum sized unsatisfiable sub-problem with size $s\leq cn,$ denoted by $%
P_{1}.$ From Lemma 1 we know that almost surely $P_{1}$ has at most $\beta
s\ln n$ constraints. Thus there exists a variable $u$ in $P_{1}$ with degree
at most $k\beta \ln n,$ i.e. the number of constraints in $P_{1}$ associated
with $u$ is not greater than $k\beta \ln n.$ Removing $u$ and the
constraints associated with $u$ from $P_{1},$ we get a sub-problem $P_{2}.$
By minimality of $P_{1},$ we know that $P_{2}$ is satisfiable, and so there
exists an assignment satisfying $P_{2}$. Suppose that the variables in $%
P_{2} $ have been assigned values by such an assignment. Now consider the
variable $u$ and the $i$ constraints associated with $u,$ where $i\leq
k\beta \ln n.$ By Definition 2 this constitutes an $i$-constraint assignment
tuple for $u,$ denoted by $T_{i,u}.$ Recall that $P_{1}$ is unsatisfiable.
This means that no value of $u$ can satisfy all the $i$ constraints. That is
to say, the variable $u$ is flawed by $T_{i,u}.$ Therefore, if a sub-problem
of size at most $cn$ is unsatisfiable, then, almost surely, there is a
variable $u$ and an $i$-constraint assignment tuple $T_{i,u}$ such that $u$
is flawed by $T_{i,u},$ where $i\leq k\beta \ln n.$ This is in contradiction
with Lemma 2 and so finishes the proof. \hfill $\Box $

\smallskip

Now we will prove that there almost surely exist a complex clause in the
refutation proofs of Model RB/RD. The complexity of a clause was defined in
[26] by Mitchell, i.e. for any refutation $\pi ,$ the complexity of a clause
$C$ in $\pi ,$ denoted by $\mu (C),$ is the size of the smallest sub-problem
$\Pi $ such that $C$ can be derived by resolution from $\phi (\Pi ).$ Along
the same line as in the proof of [26], we have the following lemma.

{\bf Lemma 4 }Let $P$ be a random CSP instance generated by Model RB/RD.
Almost surely, every refutation $\pi $ of $\phi (P)$ has a clause $C$ of
complexity $\frac{cn}{2}\leq \mu (C)\leq cn.$

{\bf Proof: }For this proof, please refer to [26]. \hfill  $\Box $

\smallskip

{\bf Lemma 5. }Let $C$ be a clause of complexity $\frac{cn}{2}\leq \mu
(C)\leq cn.$ Then, almost surely, $C$ has at least $\frac{c}{6}n$ literals,
i.e. $w(C)\geq \frac{c}{6}n$.

{\bf Proof: }We will prove this by contradiction. For a CSP instance $P,$
its CNF encoding is denoted by $\phi (P).$ Let $C$ be a clause of complexity
$\frac{cn}{2}\leq \mu (C)\leq cn$ and $P_{1}$ be the smallest problem such
that $\phi (P_{1})\models C.$ Hence, the size of $P_{1}$ is at least $\frac{c%
}{2}n$ and at most $cn$. By Lemma 1, there are at most $\beta cn\ln n$
constraints in $P_{1}.$ So, there are at most $\frac{c}{3}n$ variables with
degree greater than $3k\beta \ln n.$ Then, there are at least $\frac{c}{2}n-%
\frac{c}{3}n=\frac{c}{6}n$ variables in $P_{1}$ with degree at most $3k\beta
\ln n.$ We will prove that for these variables, almost surely, there does
not exist a variable such that no domain variable of it appears in $C.$ Now
assume that we have a variable $u$ in $P_{1}$ with degree $i\leq 3k\beta \ln
n$ and no domain variable of it appears in $C.$ Removing $u$ and the
constraints associated with it from $P_{1},$ we get a sub-problem $P_{2}.$
By minimality of $P_{1},$ we know that $\phi (P_{2})\not\models C.$ So we
can find an assignment satisfying $P_{2}$ but not satisfying $C.$ Suppose
that the propositional variables in $P_{2}$ and $C$ have been assigned
values by such an assignment. Now consider the variable $u$ and the
constraints associated with it. By Definition 2, this constitutes an $i$%
-constraint assignment tuple for $u,$ denoted by $T_{i,u}.$ By assumption,
no domain variable of $u$ appears in $C.$ So, assigning any value to $u$ \
will not affect the truth value of $C.$ Recall that $\phi (P_{1})\models C$
and $C$ is false under the current assignment. Therefore, no value of $u$
can satisfy $\phi (P_{1}),$ i.e. setting any value to $u$ will violate at
least one constraint associated with it. It follows that $u$ is flawed by $%
T_{i,u},$ i.e. there exists a flawed $i$-constraint assignment tuple with $%
i\leq 3k\beta \ln n.$ This is in contradiction with Lemma 2 and so we are
done. \hfill $\Box $

Combining Lemma 4 and Lemma 5, we have that, for a random CSP
instance $P$
generated by Model RB/RD, almost surely, $w(\phi (P)\vdash 0)\geq \frac{c}{6}%
n.$ Now, by use of Theorem 4, we finish the proof. One point worth
mentioning is that when $\alpha \geq 1,$ the initial width of clauses is
greater than or equal to the number of variables. In such a case, to make
Theorem 4 applicable, we only need to introduce some new variables and
reduce the widths of domain clauses, which has no effect on our results.

\bigskip

\noindent {\large {\bf 4. Generating Hard Satisfiable Instances}} 

\noindent As mentioned before, the finding of phase transitions in NP-complete
problems provides a good method for generating random hard instances which
are very useful in the evaluation of algorithms. In recent years, a
remarkable progress in Artificial Intelligence has been the development of
incomplete algorithms for various kinds of problems. To evaluate the
efficiency of such incomplete algorithms, we need a source to generate 
only hard satisfiable instances [3]. However, since the probability of being
satisfiable is about 0.5 at the threshold point where the hardest instances
are concentrated, the generator based on phase transitions will usually
produce a mixture of satisfiable and unsatisfiable instances. So, it is
interesting to study how the phase transition phenomenon can be used to
generate hard satisfiable instances. Besides practical importance, more
interestingly, the problem of generating random hard satisfiable instances
is related to some open problems in cryptography, e.g. computing a one-way
function, generating pseudo-random numbers and private key cryptography
[12, 21, 23].

In fact, for constraint satisfaction and Boolean satisfiability problems,
there is a natural strategy to generate instances that are guaranteed to
have at least one satisfying assignment. The strategy is as follows [3]: first
generate a random truth assignment $t,$ and then generate a certain number
of random constraints or clauses one by one to form a random instance, 
where any clause or constraint violating $t$ will
be rejected. The above strategy is very simple and can be easily
implemented. But unfortunately, this strategy was proved to be unsuitable
for random 3-SAT because it in fact produces a biased sampling of
instances with many satisfying assignments (clustered around $t$), and experiments also
show that these instances are much easier to solve than random satisfiable instances [3]. 
In the following, for convenience, we will call the satisfiable 
instances generated using the strategy as forced satisfiable instances.

Now let us look further into the problem why the strategy fails for
random 3-SAT. 
As defined in [33, 34], an {\it assignment pair} $<t_{1},t_{2}>$ is an ordered pair
of two assignments $t_{1}$ and $t_{2}.$ We say that $<t_{1},t_{2}>$ satisfies a CSP
if and only if both  $t_{1}$ and $t_{2}$ satisfy this CSP. Suppose that the
number of variables is $n$ and the domain size is $d.$ Then we have totally
$d^{n}$ possible assignments, denoted by $t_{1},t_{2},\cdots,t_{d^{n}},$ and
$d^{2n}$ possible assignment pairs. Let $t_{i}$ be a forced satisfying
assignment. Then the expected number of solutions for forced satisfiable
instances satisfying $t_{i},$ denoted by $E_{f}[N]$, is%
\[
E_{f}[N]=\frac{\overset{d^{n}}{\underset{j=1}{%
{\displaystyle\sum}
}}\Pr[<t_{i},t_{j}>]}{\Pr[<t_{i},t_{i}>]},
\]

\noindent where $\Pr[<t_{i},t_{j}>]$ denotes the probability that $<t_{i},t_{j}>$
satisfies a random instance. Note that $E_{f}[N]$ should be independent of the
choice of the forced satisfying assignment $t_{i}.$ So we have%
\[
E_{f}[N]=\frac{\underset{1\leq i,j\leq d^{n}}{%
{\displaystyle\sum}
}\Pr[<t_{i},t_{j}>]}{d^{n}\Pr[<t_{i},t_{i}>]}=\frac{E[N^{2}]}{E[N]}.
\]
\noindent where $E[N^{2}]$ and $E[N]$ are, respectively, the
second moment and the first moment of the number of solutions for instances
generated randomly. 
%It is straightforward to derive, from the results on ordered 
%pairs of assignments for random $k$-SAT in [34], that the expected number of
%solutions for random forced satisfiable instances is
%equal to $E(N^{2})/E(N),$ where $E(N^{2})$ and $E(N)$ are, respectively, the
%second moment and the first moment of the number of solutions for instances generated randomly. 
For random 3-SAT, it follows from the result on satisfying 
assignment pairs in [34] that 
asymptotically,  $E[N^{2}]$ is exponentially greater than $E^{2}[N]$. 
This conclusion can also be found in [4].
Thus, the expected number of solutions for forced satisfiable instances 
is exponentially larger than that for random satisfiable 
instances, which gives  
a good theoretical explanation of why, for random 3-SAT,
the strategy is highly biased towards generating instances with many solutions.

We now consider the problem of generating satisfiable instances for Model
RB/RD using the same strategy. Recall that when we established the exact
phase transitions for RB/RD [33], it was proved that $E[N^{2}]/E^{2}[N]$ is
asymptotically equal to 1 below the threshold, where almost all
instances are satisfiable, i.e. $E[N^{2}]/E^{2}[N]\approx 1$ for $r<r_{cr}$ 
or $p<p_{cr}$. So, we have that for
RB/RD, the expected number of solutions for forced satisfiable instances
below the threshold is asymptotically equal to that for random satisfiable 
instances, i.e. $E_{f}[N]=E[N^{2}]/E[N]\approx E[N]$. In other words, the strategy
has almost no effect on the number of solutions for RB/RD and thus 
will not lead to a biased sampling of instances with many solutions. 

In addition to the analysis above, we can also study the influence of the 
strategy on the distribution of solutions with respect to the 
forced satisfying assignment. 
%we know, from $E(N^{2})\approx E^{2}(N),$ that for randomly 
%generated instances of RB/RD, the distribution of the number of solutions is 
%quite uniform with concentration around $E(N).$  Note that the truth assignment 
%$t$ is generated randomly and the strategy will in fact generate all the possible 
%satisfiable instances with $t$ as the satisfying assignment. 
Based on the definition of {\it similarity number} in [33], we first define a 
distance on the assignments as $d^{f}(t_1,t_2)=1-S^f(\langle t_1,t_2\rangle)/n,$
where $t_1,t_2$ are two assignments, $n$ is the total number of variables and
 $S^f(\langle t_1,t_2\rangle)$ is  equal to the number of 
variables at which the two assignments take the identical values. It is easy
to see that $0\leq d^{f}(t_1,t_2)\leq 1.$ 
Let $E_{f}[X]$ and $E[X]$ respectively denote, for forced satisfiable instances 
and random satisfiable instances, the expected number of solutions with a 
fixed distance $d_{t}$ from the forced satisfying assignment.
By an analysis similar to that in [33] (pp.96-97), we have
\begin{align*}
E_{f}[X] &  =\binom{n}{nd_{t}}\left(  n^{\alpha}-1\right)  ^{nd_{t}}\frac
{\Pr[<t_{1},t_{2}>]}{\Pr[<t_{1},t_{1}>]}\text{ \ \ where }d^{f}(t_{1},t_{2}%
)=d_{t}\\
&  =\binom{n}{nd_{t}}\left(  n^{\alpha}-1\right)  ^{nd_{t}}\left[
\frac{\binom{n-nd_{t}}{k}}{\binom{n}{k}}+(1-p)\left(  1-\frac{\binom{n-nd_{t}%
}{k}}{\binom{n}{k}}\right)  \right]  ^{rn\ln n}\\
&  = \exp\left[  n\ln n\left(  r\ln\left(
1-p+p(1-d_{t})^{k}\right)  +\alpha d_{t}\right)+O(n)  \right]  .
\end{align*}
Indeed, it can be shown, from the results in [33] (pp.97-98),  
that $E_{f}[X],$ for $r<r_{cr}$ or $p<p_{cr},$
will be asymptotically maximized when $d_{t}$ takes the largest possible
value, i.e. $d_{t}=1.$
For random satisfiable instances of RB/RD, we have
\begin{align*}
E[X]  & =\binom{n}{nd_{t}}\left(  n^{\alpha}-1\right)  ^{nd_{t}}\left(
1-p\right)  ^{rn\ln n}\\
& =  \exp\left[  n\ln n\left(  r\ln(1-p)+\alpha
d_{t}\right)+O(n)  \right]  .
\end{align*}
It is straightforward to see that the same pattern holds
for this case, i.e. $E[X]$ will be asymptotically maximized when $d_{t}=1.$
So, intuitively speaking, for RB/RD, given an assignment $t,$ for both forced
satisfiable instances satisfying $t$ and random satisfiable instances, 
most solutions distribute in a place far from $t.$
This further indicates that the strategy has little effect on the distribution 
of solutions for RB/RD, and so it will not be be biased towards generating 
instances with many solutions around the forced satisfying assignment. 
For random 3-SAT, similarly, we have%
\begin{align*}
E_{f}[X]  & =\binom{n}{nd_{t}}\left[  \frac{\binom{n-nd_{t}}{3}}{\binom
{n}{3}}+\frac{6}{7}\left(  1-\frac{\binom{n-nd_{t}}{3}}{\binom{n}{3}}\right)
\right]  ^{rn}\\
& =f_{1}(n)\exp\left[  n\left(  -d_{t}\ln d_{t}-(1-d_{t})\ln(1-d_{t}%
)+r\ln\frac{6+(1-d_{t})^{3}}{7}\right)  \right]  ,
\end{align*}

\noindent and%
\begin{align*}
E[X]  & =\binom{n}{nd_{t}}\left(  \frac{7}{8}\right)  ^{rn}\\
& =f_{2}(n)\exp\left[  n\left(  -d_{t}\ln d_{t}-(1-d_{t})\ln(1-d_{t}%
)+r\ln\frac{7}{8}\right)  \right]  ,
\end{align*}
where $f_{1}(n)$ and $f_{2}(n)$ are two polynomial functions.
It follows from the results in [34] that 
as $r$ (the ratio of clauses to variables) approaches 4.25, 
$E_{f}[X]$ and $E[X]$ will be asymptotically maximized 
when $d_{t}\approx 0.24$ and $d_{t}=0.5$ respectively. This means, 
in contrast to RB/RD, that compared with random 
satisfiable instances, most solutions of forced 
satisfiable instances distribute in a place much closer to the 
forced satisfying assignment when $r$ is near the threshold.

Note that the number and the distribution of solutions are the two most
important factors determining the cost of solving satisfiable instances. 
So, we can expect, from the above analysis, that for RB/RD,
the hardness of solving forced satisfiable instances should be similar
to that of solving random satisfiable instances. 
More interestingly, it therefore seems that we can, based on the hardness
of RB/RD, propose  
a new method to generate hard satisfiable instances, i.e. generating
forced satisfiable instances of RB/RD with a large number of variables
near the threshold
identified exactly by Theorem 1 or Theorem 2.
Experimental results have further confirmed this idea\footnote{\small 
We thank Dr. Christophe Lecoutre and Liu Yang very much for performing the experiments.}. 
It is shown, in one experiment for RB with $k=2, n=30, d=15$ and $m=250,$ 
that the mean time of solving forced satisfiable instances near the
threshold is only slightly smaller (11 percent) than that of 
solving random satisfiable instances with the same 
parameters\footnote{\small As specified by the conditions of Theorem 2, 
to make exact phase transitions hold, the values of $\alpha$ and $r$ 
should not be small. So, we should choose dense CSPs with a large domain.}.  
More importantly, experiments for RB also indicate that the hardness of 
solving forced satisfiable 
instances grows exponentially with the number of variables\footnote
{\small According to the
definitions of RB/RD and Theorems 1 and 2, the parameters $\alpha,$ $r$ and $p$ 
should be fixed when $n$ increases. The values of the threshold points can also
be obtained from these two theorems.}
near the threshold, 
and we can, in fact, generate forced satisfiable instances appearing to be
very hard to solve (for both complete and incomplete algorithms) even when the 
number of variables is only moderately 
large (e.g. $k=2, n=59, \alpha=0.8$ and $r=0.8/\ln\frac{4}{3}$ with 
constraint tightness 
$p=p_{cr}=0.25$ computed by Theorem 2, or equivalently expressed
as $k=2, n=59, d=26$ and $m=669$ with the same tightness\footnote
{\small If non-integer values occur in the computation 
of $d$ and $m$ from $n,$ $\alpha$ and $r,$ then we round them to the 
nearest integers.})\footnote 
{\small Benchmarks of Model RB (in both SAT and CSP format) are available at
www.nlsde.buaa.edu.cn/\symbol{126}kexu/ benchmarks/benchmarks.htm.}.
%More interestingly, we have successfully generated some forced 
%satisfiable instances which appear to be very hard to solve, i.e. these instances can not be
%solved by state-of-the-art CSP algorithms in a reasonable time (e.g. 1 day).
Although there have been some other ways to generate hard satisfiable 
instances empirically, e.g. the quasigroup method [3], we think that 
the simple and natural method presented in this paper, 
based on models (i.e. Model RB/RD) with exact phase transitions and many 
hard instances, should be well worth further investigation.  

\bigskip
\smallskip
\noindent {\large {\bf 5. Exponential Lower Bounds for Satisfiable Instances of Model RB/RD}} 

\noindent For random CSP instances of RB/RD, we know from Theorems 1 and 2 that almost
surely, they are satisfiable below the threshold and unsatisfiable above the
threshold. For satisfiable instances, there are no resolution proofs, or, if
any, the resolution proofs are of infinite length. Therefore, the exponential
resolution lower bounds, established in Theorem 4, are of interest only for
instances above the threshold. Also, in many other cases, exponential lower
bounds have been shown only for unsatisfiable instances, and it seems quite
difficult to derive such lower bounds for satisfiable instances. A recent progress
in this direction, made by Achlioptas et. al. [5], is that 
exponential lower bounds have been established for certain natural
DPLL algorithms on some provably satisfiable instances of random $k$-SAT for $k\geq 4.$
In this section, we will analyze the complexity of solving RB/RD below the threshold
and obtain the following results.

{\bf Theorem 5} \ Given a random CSP instance of RB/RD with $r_{cr}-\epsilon
_{r}<r\leq r_{cr}$ or $p_{cr}-\epsilon_{p}<p\leq p_{cr}$, where $\epsilon_{r}%
=-\frac{\alpha}{\ln(1-p)}+\frac{\alpha(1-\frac{c}{24})}{\ln\left(  1-p\left(
1-\frac{c^{k}}{12^{k}}\right)  \right)  }$ and $\epsilon_{p}=\left[
1-\exp\left(  -\frac{\alpha}{r}(1-\frac{c}{24})\right)  \right]  \frac{12^{k}%
}{12^{k}-c^{k}}-1+\exp\left(  -\frac{\alpha}{r}\right)  $ are two positive constants, we uniformly select
without repetition $\frac{c}{12}n$ variables, and assign each of these
variables a value from its domain at random. If such values does not violate any constraint, 
then, almost surely, the residual
formula is unsatisfiable and has no tree-like resolution proofs of less than
exponential size.

{\bf Proof:} Let $E[X]$ denote the expected number of assignments satisfying the
residual formula. By assumption, the partial assignment to
the $\frac{c}{12}n$ variables does not violate any constraint. Then%
\[
E[X]= d^{n-\frac{c}{12}n}\left[  1-p\left(  1-\frac{c^{k}}{12^{k}}\right)
\right]  ^{rn\ln n}.
\]

\noindent For $r_{cr}-\epsilon_{r}<r\leq r_{cr},$ we have%
\begin{align*}
E[X] &  \leq n^{\alpha n(1-\frac{c}{12})}\left[  1-p\left(  1-\frac{c^{k}%
}{12^{k}}\right)  \right]  ^{(r_{cr}-\epsilon_{r})n\ln n}\\
&  \leq\exp\left[  \left(  -\epsilon_{r}\ln\left(  1-p\left(  1-\frac{c^{k}%
}{12^{k}}\right)  \right)  -\frac{\alpha c}{12}\right)  n\ln n\right]  \\
&  =\exp\left(  -\frac{\alpha c}{24}n\ln n\right)  =o(1).
\end{align*}

\noindent By Markov's inequality, we know that the residual formula will be almost
surely unsatisfiable. For the phase transition with respect to $p,$ the proof
can be done similarly. 
Now we prove that for the residual formula, any sub-problem of size at most
$cn$ is almost surely satisfiable. Based on the
proofs of Lemmas 2 and 3, we only need to show that for any sub-problem with
size $1\leq s\leq$ $cn$ containing unassigned variables, there almost surely
exists an unassigned variable with degree at most $3k\beta\ln n.$ Thus, it is
sufficient to prove that for any sub-problem with size $1+\frac{c}{12}n\leq
s\leq$ $cn$ $+\frac{c}{12}n$ containing the $\frac{c}{12}n$ assigned
variables, there almost surely exists an unassigned variable with degree at
most $3k\beta\ln n.$ For such a sub-problem, the probability that an
unassigned variable has a degree at least $3k\beta\ln n$ is not greater than%
\[
\binom{rn\ln n}{b}\binom{kb}{b}\left(  \frac{1}{n}\right)  ^{b}\left(
\frac{s}{n}\right)  ^{kb-b}\text{ \ where }b=3k\beta\ln n.
\]

\noindent Then, the probabilty that all the unassigned variables have degrees at least
$3k\beta\ln n$ is not greater than%
\[
\left[  \binom{rn\ln n}{b}\binom{kb}{b}\left(  \frac{1}{n}\right)  ^{b}\left(
\frac{s}{n}\right)  ^{kb-b}\right]  ^{s-\frac{c}{12}n}.
\]

\noindent There are $\binom{n-\frac{c}{12}n}{s-\frac{c}{12}n}$ possible choices for such
sub-problems$.$ So the expected number of such sub-problems with size
$1+\frac{c}{12}n\leq s\leq$ $cn$ $+\frac{c}{12}n$ is at most %

\begin{align*}
& \underset{s=1+\frac{c}{12}n}{\overset{cn+\frac{c}{12}n}{%
%TCIMACRO{\dsum }%
%BeginExpansion
{\displaystyle\sum}
%EndExpansion
}}\binom{n-\frac{c}{12}n}{s-\frac{c}{12}n}\left[  \binom{rn\ln n}{b}\binom
{kb}{b}\left(  \frac{1}{n}\right)  ^{b}\left(  \frac{s}{n}\right)
^{kb-b}\right]  ^{s-\frac{c}{12}n}\text{ where }b=3k\beta\ln n\\
& \leq\underset{s=1+\frac{c}{12}n}{\overset{cn+\frac{c}{12}n}{%
%TCIMACRO{\dsum }%
%BeginExpansion
{\displaystyle\sum}
%EndExpansion
}}\left(  \frac{e(n-\frac{c}{12}n)}{s-\frac{c}{12}n}\right)  ^{s-\frac{c}%
{12}n}\left[  \left(  \frac{rn\ln n}{b}\right)  ^{b}\left(  \frac{ekb}%
{b}\right)  ^{b}\left(  \frac{1}{n}\right)  ^{b}\left(  \frac{s}{n}\right)
^{kb-b}\right]  ^{s-\frac{c}{12}n}\\
& \leq\underset{s=1+\frac{c}{12}n}{\overset{cn+\frac{c}{12}n}{%
%TCIMACRO{\dsum }%
%BeginExpansion
{\displaystyle\sum}
%EndExpansion
}}\left[  en\left(  \frac{re}{3\beta}\right)  ^{3k\beta\ln n}\left(  \frac
{s}{n}\right)  ^{3k(k-1)\beta\ln n}\right]  ^{s-\frac{c}{12}n}.
\end{align*}

\noindent In the proof of Lemma 1, we define $e\left(  \frac{re}{\beta
}\right)  ^{\beta\ln n}<n^{c_{1}}$ and $c<\frac{1}{2}\exp\left(
-\frac{2+c_{1}}{(k-1)\beta}\right)  .$ Substituting them into the above
inequality, we get%

\begin{align*}
& \underset{s=1+\frac{c}{12}n}{\overset{cn+\frac{c}{12}n}{%
%TCIMACRO{\dsum }%
%BeginExpansion
{\displaystyle\sum}
%EndExpansion
}}\left[  en\left(  \frac{re}{3\beta}\right)  ^{3k\beta\ln n}\left(  \frac
{s}{n}\right)  ^{3k(k-1)\beta\ln n}\right]  ^{s-\frac{c}{12}n}\text{ where
}1+\frac{c}{12}n\leq s\leq cn+\frac{c}{12}n\text{\  }\\
& \leq\underset{s=1+\frac{c}{12}n}{\overset{cn+\frac{c}{12}n}{%
%TCIMACRO{\dsum }%
%BeginExpansion
{\displaystyle\sum}
%EndExpansion
}}\left[  en\frac{n^{3kc_{1}}}{e^{3k}}\frac{1}{3^{3k\beta\ln n}}%
n^{-3kc_{1}-6k}\right]  \\
& =\underset{s=1+\frac{c}{12}n}{\overset{cn+\frac{c}{12}n}{%
%TCIMACRO{\dsum }%
%BeginExpansion
{\displaystyle\sum}
%EndExpansion
}}O\left(  \frac{1}{n^{2}}\right)  =o(1),
\end{align*}
as required. Now for the residual formula, Lemmas 3 and 4 follow immediately.
Recall that in Lemma 5, we prove that there are at
least $\frac{c}{6}n$ variables in $P_{1}$ with degree at most $3k\beta\ln n.$
For the residual formula where $\frac{c}{12}n$ variables have been assigned
values, there are at least $\frac{c}{12}n$ variables in $P_{1}$ with degree at
most $3k\beta\ln n$. Similarly, we can prove that almost surely, there is a
clause with at least $\frac{c}{12}n$ literals for the residual formula. By
Theorem 4, we finish the proof. Note that the constant $c$ can be
chosen to monotonically decrease with $r$ or $p.$ Here we can, therefore,
take the value of $c$ as that for $r=r_{cr}$ or $p=p_{cr}$ and try to
make it as small as possible (in order to guarantee that $\epsilon_{r}$ and $\epsilon_{p}$
are two positive constants). \hfill $\Box $

Generally speaking, different search algorithms use different strategies to
search for solutions. Rather than focusing on some specific algorithms, we relate
the hardness of solving satisfiable instances to that of solving unsatisfiable
sub-problems, because if it takes a long time to solve the sub-problems
generated in the search process, then the original problem can not be solved
quickly [24]. Theorem 5 indicates that for satisfiable instances of RB/RD below
and close to the threshold, if a resolution-based algorithm can not detect any 
contradiction
in the early stage of a search branch, then the algorithm will, very likely, 
generate a large-sized unsatisfiable sub-problem. As a result, it will, then,
almost surely take exponential time to explore large subtrees to prove the
unsatisfiability of the sub-problem. 
Indeed, there are exponentially many large-sized unsatisfiable sub-problems. 
More precisely, it can be computed
that the total number of residual formulas with $\frac{c}{12}n$ assigned
variables and without violating any constraint is at least%
\begin{align*}
\binom{n}{\frac{c}{12}n}d^{\frac{c}{12}n}\left(  1-(\frac{c}{12})^{k}p\right)
^{r_{cr}n\ln n}  & \geq\binom{n}{\frac{c}{12}n}\exp\left[  \frac{\alpha cn\ln
n}{12}\left(  1-\frac{p}{12\ln(1-p)}\right)  \right]  \\
& =\exp\left(  \Omega(n\ln n)\right)  .
\end{align*}
So, intuitively speaking, when solving
satisfiable instances of RB/RD near the threshold, backtrack-style algorithms
will very easily fall into pitfalls with no solutions, and then, worse still,
take a long time to escape from these pitfalls. To our best knowledge, this is
the first result on the complexity of solving satisfiable instances near the proved
threshold, which can help us to gain a better understanding of the extreme
hardness of instances in the phase transition region.

For random forced satisfiable instances near the proved threshold, similarly, we have
the following result.

{\bf Theorem 6} \ Given a random forced satisfiable instance of RB/RD with
$r_{cr}-\epsilon_{r}<r\leq r_{cr}$ or $p_{cr}-\epsilon_{p}<p\leq p_{cr}$, where $\epsilon_{r}%
=-\frac{\alpha}{\ln(1-p)}+\frac{\alpha(1-\frac{c}{24})}{\ln\left(  1-p\left(
1-\frac{c^{k}}{12^{k}}\right)  \right)  }$ and $\epsilon_{p}=\left[
1-\exp\left(  -\frac{\alpha}{r}(1-\frac{c}{24})\right)  \right]  \frac{12^{k}%
}{12^{k}-c^{k}}-1+\exp\left(  -\frac{\alpha}{r}\right)  $ are two positive constants, we
uniformly select without repetition $\frac{c}{12}n$ variables, and assign each
of these variables a value from its domain at random. If such values does not violate 
any constraint, then, almost surely, the
residual formula is unsatisfiable and has no tree-like resolution proofs of
less than exponential size.

{\bf Proof:} Due to limited space, we only give the proof for the case of the phase
transition with respect to $r$ in Model RD with $\frac{1}{k}<\alpha<1. $ The
other cases can be handled similarly. Assume that we have two assignments
$t_{1}$ and $t_{2}$ and the similarity number [33] between $t_{1}$ and $t_{2}$
is $S^{f}(<t_{1},t_{2}>)=S.$ Let $P$ be a random instance of Model RD. Based
on the analysis in [33] (p.96), the probability that both $t_{1}$ and $t_{2}$ satisfy
$P$ is%
\[
\Pr[t_{1}\text{ and }t_{2}\text{ satisfy }P]=\left[  (1-p)\frac{\binom{S}{k}%
}{\binom{n}{k}}+(1-p)^{2}\left(  1-\frac{\binom{S}{k}}{\binom{n}{k}}\right)
\right]  ^{rn\ln n}.
\]

\noindent Now we suppose that $t_{0}$ is a random forced satisfying assignment and $t$
is an assignment with $S^{f}(<t_{0},t>)=S.$ Let $P_{sat}$ be a random forced
satisfiable formula of Model RD with $t_{0}$ as the forced satisfying
assignment. Then the probability that $t$ satisfies $P_{sat}$ is%
\begin{align*}
\Pr[t\text{ satisfies }P_{sat}]  & =\frac{\Pr[t_{0}\text{ and }t\text{ satisfy
}P]}{\Pr[t_{0}\text{ satisfy }P]}\\
& =\left[  1-p+p\left(  \left(  \frac{S}{n}\right)  ^{k}+\frac{g\left(
\frac{S}{n}\right)  }{n}\right)  +O\left(  \frac{1}{n^{2}}\right)  \right]
^{rn\ln n}.
\end{align*}

\noindent where $g(s)=\frac{k(k-1)}{2}(s^{k}-s^{k-1}).$ Now, for the random forced
satisfiable formula $P_{sat},$ we uniformly select without repetition
$\frac{c}{12}n$ variables and then assign each of these variables a value from
its domain at random. By the standard Chernoff bound, it is easy to show
that the similarity number between the forced satisfying assignment $t_{0}$
and the random partial assignment to the $\frac{c}{12}n$ variables is almost
surely less than $\frac{c}{6}n^{1-\alpha}.$ For the residual formula, we have
totally $d^{n-\frac{c}{12}n}$ possible assignments. Let $t^{\prime}$ be an
assignment to the $n-\frac{c}{12}n$ variables of the residual formula with
$S^{f}(<t_{0},t^{\prime}>)=S^{\prime}.$ By assumption, the partial assignment to
the $\frac{c}{12}n$ variables does not violate any constraint. 
Thus, almost surely, the probability
that $t^{\prime}$ satisfies the residual formula is at most%
\[
\left[  1-p\left(  1-\frac{c^{k}}{12^{k}}\right)  \left(  1-\left(  \frac
{c}{6n^{\alpha}}+\frac{S^{\prime}}{n}\right)  ^{k}O(1)-\frac{g\left(  \frac
{c}{6n^{\alpha}}+\frac{S^{\prime}}{n}\right)  }{n}O(1)\right)  \right]  ^{rn\ln
n}.
\]

\noindent Let $E[X]$ be the expected number of assignments satisfying the residual
formula. Similar to the asymptotic analysis in [33] (p.99), for
$r_{cr}-\epsilon_{r}<r\leq r_{cr},$ we have%
\begin{align*}
E[X] &  \leq\overset{n-\frac{c}{12}n}{\underset{S^{\prime}=0}%
{{\displaystyle\sum}}}\binom{n-\frac{c}{12}n}{S^{\prime}}\left(  n^{\alpha
}-1\right)  ^{n-\frac{c}{12}n-S^{\prime}}\\
&  \cdot\left[  1-p\left(  1-\frac{c^{k}}{12^{k}}\right)  \left(  1-\left(
\frac{c}{6n^{\alpha}}+\frac{S^{\prime}}{n}\right)  ^{k}O(1)-\frac{g\left(
\frac{c}{6n^{\alpha}}+\frac{S^{\prime}}{n}\right)  }{n}O(1)\right)  \right]
^{rn\ln n}\\
&  \approx n^{\alpha n(1-\frac{c}{12})}\left[  1-p\left(  1-\frac{c^{k}%
}{12^{k}}\right)  \right]  ^{rn\ln n}\underset{S^{\prime}=0}%
{{\displaystyle\sum}}\binom{n-\frac{c}{12}n}{S^{\prime}}\left(  \frac
{1}{n^{\alpha}}\right)  ^{S^{\prime}}\left(  1-\frac{1}{n^{\alpha}}\right)
^{n-S^{\prime}}\text{ \ for }\frac{1}{k}<\alpha<1\\
&  \approx n^{\alpha n(1-\frac{c}{12})}\left[  1-p\left(  1-\frac{c^{k}%
}{12^{k}}\right)  \right]  ^{rn\ln n}.
\end{align*}

\noindent Note that the forced satisfying assignment has no effect on the
structure of constraint graphs.
The rest of the proof is identical to that in Theorem 5 and so we are done. \hfill $\Box $

The above theorem, as far as we know, is the first complexity result of
resolution-based algorithms on forced satisfiable instances, which further 
provides, from another aspect, a
strong theoretical support for the method of generating hard satisfiable
instances proposed in the last section.

\bigskip

\noindent {\large {\bf 6. Conclusions}}

\smallskip

\noindent In this paper, by encoding CSPs into CNF formulas, we proved
exponential lower bounds for tree-like resolution proofs of two random CSP
models with exact phase transitions, i.e. Model RB/RD. This result suggests
that we not only introduce new families of CNF formulas hard for resolution,
which is a central task of Proof-Complexity theory, but also propose models 
with both many hard instances and exact phase transitions. More interestingly, 
it is shown both theoretically and experimentally that an application of RB/RD 
might be in the generation of hard satisfiable instances, which is further
supported by the exponential lower bounds established in Section 6.

As mentioned before, there are some other NP-complete problems with proved
exact phase transitions, e.g. Hamiltonian cycle problem and random 2+$p$-SAT
($0<p\leq 0.4$). However, it has been shown either experimentally or
theoretically that the instances produced by these problems are generally
easy to solve. So one would naturally ask what the main difference between
these ``easy" NP-complete problems and RB/RD is. It seems that for these ``easy"
NP-complete problems with exact phase transitions, they usually have some 
kind of local property which can be used to design polynomial time algorithms 
working with high probability, and the exact phase transitions are, in fact, 
obtained by probabilistic analysis of such algorithms.
So, it appears that if a problem has exact phase transitions obtained
by algorithm analysis, then it also means that the problem is  
not hard to solve. For RB/RD, the situation is, however, completely different. 
More specifically, the exact phase transitions of RB/RD are 
obtained, not by analysis of algorithms, but by use of the 
first and the second moment methods which say nothing about the local 
property of the problem and are, therefore, unlikely to be useful for designing
more efficient algorithms. 
Thus, it seems that RB/RD, unlike the ``easy" NP-complete problems, 
can indeed provide a reliable source 
to generate random benchmark instances, as many and as hard as we need.

Note that more recently, Frieze and Wormald [15] studied random $k$-SAT for moderately
growing $k,$ i.e. $k=k(n)$ satisfies $k-\log _{2}n\rightarrow \infty$ 
where $n$ is the number of variables. 
For this model, they established similarly, by use of the first and the second 
moment methods, that there exists a satisfiability threshold at
which the number of clauses is $m=2^{k}n\ln 2$. 
%They proved that for this model, a random instance is satisfiable (unsatisfiable)
%with probability tending to 1 as the number of variables $n\rightarrow\infty$ 
%if the number of clauses $m\leq (1-\epsilon)m_0$ 
%($m\geq (1-\epsilon)m_0$) where $m_0=2^{k}n\ln2$ and $\epsilon=\epsilon(n)>0$
%satisfying $\epsilon n\rightarrow\infty.$
From Beame et al's earlier work on the complexity of unsatisfiability 
proofs for random $k$-SAT formulas [6, 7], we know that the 
size of resolution refutations for this
model is exponential with high probability. So, the variant of
random $k$-SAT studied by Frieze and Wormald is also a model with both proved
phase transitions and many hard instances. 
%But unlike the phase transitions
%of random $k$-SAT with fixed $k$ such as random 3-SAT, the critical value of 
%the ratio of clauses to variables for this variant model is not a
%constant but grows with the number of variables. 

To gain a better understanding of Model RB/RD, we now 
make a comparison of them with the well-studied
%Now, we can also make a comparison between Model RB/RD and the well-studied 
random 3-SAT of similar proof complexity. 
First, we think that the exact phase transitions should be one advantage 
of RB/RD, which 
can help us to locate the hardest instances more 
precisely and conveniently when implementing 
large-scale computational experiments. As for the theoretical aspect, it seems
that RB/RD, intrinsically, are much mathematically 
easier to analyze 
than random 3-SAT, such as in the derivation of thresholds. 
From a personal perspective, we think that 
such mathematical tractability should be another advantage of RB/RD, making 
it possible to obtain some interesting results which do not hold or can not 
be easily obtained for random 3-SAT, just as shown on forced satisfiable
instances.

In summary, the Hamiltonian cycle problem, random 3-SAT and Model RB/RD,
respectively, exhibit three different kinds of phase transition behavior in
NP-complete problems. Compared with the former two that have been
extensively explored in the past decade, the third one (i.e. the phase
transition behavior with both exact thresholds and many hard instances),
due to various reasons, has not received much attention so far. 
From this point, the main contribution of this paper, we can say, is not 
in the mathematical techniques used, nor the concrete models studied 
(although such models are useful for CSP research in their own right), but  
pointing out an interesting behavior for study.  
Finally, we hope that more investigations, either experimental or
theoretical, will be carried out on this behavior, and we also believe that
such studies will lead to deep insights and new discoveries in this active
area of research (i.e. on phase transitions and computational complexity). 

\bigskip

\noindent {\large {\bf References}}
{\small
%\smallskip

\begin{enumerate}
\item D. Achlioptas, L. Kirousis, E. Kranakis and D. Krizanc, Rigorous
results for random (2+$p$)-SAT, In: {\it Proceedings of RALCOM-97}, pp.1-10.

\item D. Achlioptas, LM Kirousis, E. Kranakis, D. Krizanc, M. SO Molloy, and YC. Stamatiou, 
Random Constraint
Satisfaction: A More Accurate Picture, In: {\it Proc. Third International Conference on Principles and
Practice of Constraint Programming} (CP 97), LNCS 1330, pp.107-120, 1997. 

\item D. Achlioptas, C. Gomes, H. Kautz, and B. Selman, Generating Satisfiable 
Problem Instances, In: {\it Proceedings of AAAI-00}, pp.256-301.

\item D. Achlioptas and C. Moore. The Asymptotic Order of the Random $k$-SAT Threshold.
In {\it Proc. FOCS 2002}, pp.779-788.

\item D. Achlioptas, P. Beame and M. Molloy. Exponential Bounds for DPLL below the 
Satisfiability Threshold. In: {\it Proc. SODA 2004}, to appear.

\item P. Beame, R. Karp, T. Pitassi, and M. Saks. On the complexity of 
unsatisfiability proofs for random $k$-CNF formulas. In: {\it Proceeding of STOC-98}, pp.561-571.

\item P. Beame, R. Karp, T. Pitassi, and M. Saks. The efficiency of resolution 
and Davis-Putnam procedures. {\it SIAM Journal on Computing}, 31(4):1048-1075, 2002.

\item E. Ben-Sasson and A. Wigderson. Short proofs are narrow - resolution
made simple. {\it Journal of the ACM}, 48(2):149-169, 2001.

\item B. Bollob\'{a}s, T.I. Fenner and A.M. Frieze. An algorithm for finding
Hamilton paths and cycles in random graphs. {\it Combinatorica}
7(4):327-341, 1987.

\item V. Chv\'{a}tal and E. Szemer\'{e}di. Many hard examples for
resolution. {\it Journal of the ACM}, 35(4) (1988) 759-208.

\item V. Chv\'{a}tal and B. Reed. Miks gets some (the odds are on his side).
In: {\it Proceedings of the 33rd IEEE Symp. on Foundations of Computer
Science}, pages 620-627, 1992.

\item S. Cook and D. Mitchell. Finding Hard Instances of the Satisfiability
Problem: A Survey, In: {\it Satisfiability Problem: Theory and Applications}%
. Du, Gu and Pardalos (Eds). DIMACS Series in Discrete Mathematics and
Theoretical Computer Science, Volume 35, 1997.

\item O. Dubois and J. Mandler. The 3-XORSAT threshold. In: {\it Proc. FOCS 2002}.

\item E. Friedgut, Sharp thresholds of graph properties, and the k-sat
problem. With an appendix by Jean Bourgain. {\it Journal of the American
Mathematical Society} 12 (1999) 1017-1054.

\item  A.M. Frieze and N.C. Wormald. Random $k$-SAT: A tight threshold for moderately 
growing $k,$ In: {\it Proceedings of the Fifth International Symposium on Theory 
and Applications of Satisfiability Testing}, pp.1-6, 2002.

\item A. Flaxman. A sharp threshold for a random constraint satisfaction problem, preprint.

\item A. Frieze and M. Molloy. The satisfiability threshold for randomly generated 
binary constraint satisfaction problems. In: {\it Proceedings of RANDOM-03}, 2003.

\item Y. Gao and J. Culberson. Resolution Complexity of Random Constraint Satisfaction 
Problems: Another Half of the Story. In: {\it Proc. of LICS-03, Workshop on Typical Case 
Complexity and Phase Transitions}, Ottawa, Canada, June, 2003.

\item I.P. Gent, E. MacIntyre, P. Prosser, B.M. Smith and T. Walsh, Random Constraint 
Satisfaction: flaws and structures. {\it Journal of Constraints} 6(4), 345-372, 2001.

\item A. Goerdt. A threshold for unsatisfiability. In: {\it 17th
International Symposium of Mathematical Foundations of Computer Science},
Springer LNCS 629 (1992), pp.264-275.

\item R. Impagliazzo, L. Levin, and M. Luby. Pseudo-random number generation from 
one-way functions. In: {\it Proceedings of STOC-89}, pp.12-24. 

\item M. Koml\'{o}s and E. Szemer\'{e}di. Limit distribution for the
existence of a Hamilton cycle in a random graph. {\it Discrete Mathematics},
43, pp.55-63, 1983.

\item M. Luby. Pseudorandomness and Cryptographic Applications. Princeton 
University Press, 1996. 

\item D. Mitchell: Hard Problems for CSP Algorithms. In: {\it Proceedings of 15th
National Conf. on Artificial Intelligence} (AAAI-98), pp.398-405, 1998.

\item D. Mitchell, B. Selman, and H. Levesque. Hard and easy distributions
of sat problems. In: {\it Proceedings of 10th National Conf. on Artificial
Intelligence} (AAAI-92), pp.459-465, 1992.

\item D. Mitchell. Resolution Complexity of Random Constraints, In: {\it %
Proceedings of CP 2002}, LNCS 2470, pp.295-309. 

\item M. Molloy. Models for Random Constraint Satisfaction Problems,
submitted. Conference version in {\it Proceedings of STOC 2002}.

\item M. Molloy and M. Salavatipour. The resolution complexity of random 
constraint satisfaction problems. In: {\it Proc. FOCS-03}, 2003.

\item R. Monasson, R. Zecchina, S. Kirkpatrick, B. Selman and L. Troyansky.
Determining computational complexity from characteristic phase transitions.
{\it Nature}, 400(8):133-137, 1999.

\item R. Monasson, R. Zecchina, S. Kirkpatrick, B. Selman and L. Troyansky,
Phase transition and search Cost in the 2+$p$-SAT problem, In: {\it 4th
Workshop on Physics and Computation}, Boston University 22-24 November 1996,
(PhysComp96).

\item B.M. Smith. Constructing an Asymptotic Phase Transition in Random Binary 
Constraint Satisfaction Problems. {\it Theoretical Computer Science}, vol. 265, 
pp. 265-283 (Special Issue on NP-Hardness and Phase Transitions), 2001.

\item B. Vandegriend and J. Culberson. The $G_{n,m}$ phase transition is not
hard for the Hamiltonian Cycle problem. {\it Journal of Artificial
Intelligence Research}, 9:219-245, 1998.

\item K. Xu and W. Li. Exact Phase Transitions in Random Constraint
Satisfaction Problems. {\it Journal of Artificial Intelligence Research},
12:93-103, 2000. 

\item K. Xu. A Study on the Phase Transitions of SAT and CSP (in Chinese). 
Ph.D. Thesis, Beihang University, 2000.

\item K. Xu and W. Li. On the Average Similarity Degree between Solutions of Random 
$k$-SAT and Random CSPs. {\it Discrete Applied Mathematics}, to appear. 

\medskip
\end{enumerate}
}

\noindent {\large {\bf Appendix}}

\smallskip

Now we consider the proof of Lemma 2 for Model RB. Given a variable $u$ an $i
$-constraint assignment tuple $T_{i,u}.$ It is easy to see that the
probability that $u$ is flawed by $T_{i,u}$ increases with the number of
constraints $i.$ Thus we have

\[
\Pr (T_{i,u}\text{ is flawed})|_{i\leq 3k\beta \ln n}\leq \Pr (T_{i,u}\text{
is flawed})|_{i=3k\beta \ln n}.
\]

\noindent For the variable $u,$ there are $d=n^{\alpha }$ values in its
domain, denoted by $v_{1},v_{2},\cdots ,v_{d}.$ Let $\Pr (A_{j})$ denote the
probability that $v_{j}$ is not flawed by $T_{i,u}.$ Thus the probability
that at least one value is not flawed by $T_{i,u},$ i.e. the probability
that the variable $u$ is not flawed by $T_{i,u}$ is
\begin{eqnarray*}
\Pr (A_{1}\cup A_{2}\cup \cdots \cup A_{d}) &=&\underset{1\leq p\leq d}{\sum
}\Pr (A_{p})-\underset{1\leq p,q\leq d,p\neq q}{\sum }\Pr (A_{p}A_{q}) \\
&&+\cdots +(-1)^{d-1}\Pr (A_{1}A_{2}\cdots A_{d}).
\end{eqnarray*}

\noindent Then
\begin{eqnarray*}
\Pr (T_{i,u}\text{ is flawed}) &=&1-\Pr (A_{1}\cup A_{2}\cup \cdots \cup
A_{d}) \\
&=&1+\underset{j=1}{\overset{d}{\sum }}(-1)^{j}\binom{d}{j}\Pr
(A_{1}A_{2}\cdots A_{j}).
\end{eqnarray*}

\noindent Recall that in Model RB, for each constraint, we uniformly select
without repetition $pd^{k}$ incompatible tuples of values and each
constraint is generated independently. So we have
\begin{eqnarray*}
\Pr (A_{1}A_{2}\cdots A_{j}) &=&\left[ \frac{\binom{d^{k}-j}{pd^{k}}}{\binom{%
d^{k}}{pd^{k}}}\right] ^{i} \\
&=&\left[ \frac{(d^{k}-pd^{k})(d^{k}-pd^{k}-1)\cdots (d^{k}-pd^{k}-j+1)}{%
d^{k}(d^{k}-1)\cdots (d^{k}-j+1)}\right] ^{i}.
\end{eqnarray*}

\noindent Note that $j\leq d=n^{\alpha }$ and $k\geq 2.$ Now consider the
case of $i=3k\beta \ln n,$ where $\beta =\frac{\alpha }{6k\ln \frac{1}{1-p}}%
. $ By asymptotic analysis, we have
\begin{eqnarray*}
&&\Pr (A_{1}A_{2}\cdots A_{j})|_{i=3k\beta \ln n} \\
&=&[(1-p)(\frac{1-p-\frac{1}{n^{k\alpha }}}{1-\frac{1}{n^{k\alpha }}})(\frac{%
1-p-\frac{2}{n^{k\alpha }}}{1-\frac{2}{n^{k\alpha }}})\cdots (\frac{1-p-%
\frac{j-1}{n^{k\alpha }}}{1-\frac{j-1}{n^{k\alpha }}})]^{3k\beta \ln n} \\
&=&[(1-p)^{3k\beta \ln n}]^{j}[1-\frac{p}{1-p}\frac{(j-1)j}{2n^{k\alpha }}+O(%
\frac{j^{4}}{n^{2k\alpha }})]^{3k\beta \ln n} \\
&=&(n^{-\frac{\alpha }{2}})^{j}[1-\frac{p}{1-p}\frac{(j-1)j}{2n^{k\alpha }}%
+O(\frac{j^{4}}{n^{2k\alpha }})]^{3k\beta \ln n}.
\end{eqnarray*}

\noindent Let $H(j)=[1-\frac{p}{1-p}\frac{(j-1)j}{2n^{k\alpha }}+O(\frac{%
j^{4}}{n^{2k\alpha }})]^{3k\beta \ln n}.$ Then we get
\begin{eqnarray*}
\Pr (T_{i,u}\text{ is flawed})|_{i=3k\beta \ln n} &=&1+\underset{j=1}{%
\overset{n^{\alpha }}{\sum }}(-1)^{j}\binom{n^{\alpha }}{j}\Pr
(A_{1}A_{2}\cdots A_{j})|_{i=3k\beta \ln n} \\
&=&1+\underset{j=1}{\overset{n^{\alpha }}{\sum }}(-1)^{j}\binom{n^{\alpha }}{%
j}(n^{-\frac{\alpha }{2}})^{j}H(j).
\end{eqnarray*}

\noindent For $0\leq j\leq n^{\frac{4}{5}\alpha },$ we can easily show that $%
H(j)=1+o(1).$ Therefore,
\begin{eqnarray*}
&&\Pr (T_{i,u}\text{ is flawed})|_{i=3k\beta \ln n} \\
&\approx &1+\underset{j=1}{\overset{n^{\alpha }}{\sum }}(-1)^{j}\binom{%
n^{\alpha }}{j}(n^{-\frac{\alpha }{2}})^{j}+\underset{j=n^{\frac{4}{5}\alpha
}}{\overset{n^{\alpha }}{\sum }}(-1)^{j}\binom{n^{\alpha }}{j}(n^{-\frac{%
\alpha }{2}})^{j}(H(j)-1) \\
&=&(1-\frac{1}{n^{\frac{\alpha }{2}}})^{n^{\alpha }}+\underset{j=n^{\frac{4}{%
5}\alpha }}{\overset{n^{\alpha }}{\sum }}(-1)^{j}\binom{n^{\alpha }}{j}(n^{-%
\frac{\alpha }{2}})^{j}(H(j)-1) \\
&\approx &e^{-n^{\frac{\alpha }{2}}}+\underset{j=n^{\frac{4}{5}\alpha }}{%
\overset{n^{\alpha }}{\sum }}(-1)^{j}\binom{n^{\alpha }}{j}(n^{-\frac{\alpha
}{2}})^{j}(H(j)-1).
\end{eqnarray*}

\noindent It is easy to verify that

\[
\binom{n^{\alpha }}{j}(n^{-\frac{\alpha }{2}})^{j}\leq (\frac{en^{\alpha }}{j%
})^{j}(n^{-\frac{\alpha }{2}})^{j}=e^{j-j\ln j+\frac{\alpha }{2}j\ln n}.
\]

\noindent Let $B(j)=j-j\ln j+\frac{\alpha }{2}j\ln n.$ Differentiating $B(j)$
with respect to $j,$ we obtain

\[
B^{\prime }(j)=\frac{\alpha }{2}\ln n-\ln j<0\text{ when }j\geq n^{\frac{4}{5%
}\alpha }.
\]

\noindent So for $n^{\frac{4}{5}\alpha }\leq j\leq n^{\alpha },$ we have

\[
\binom{n^{\alpha }}{j}(n^{-\frac{\alpha }{2}})^{j}\leq e^{B(n^{\frac{4}{5}%
\alpha })}=(\frac{e}{n^{\frac{3}{10}\alpha }})^{n^{\frac{4}{5}\alpha
}}=o(e^{-n^{\frac{4}{5}\alpha }}).
\]

\noindent Note that $H(j)=O(n^{c_{2}})$ for $n^{\frac{4}{5}\alpha }\leq
j\leq n^{\alpha },$ where $c_{2}>0$ is a constant. Hence,
\begin{eqnarray*}
|\underset{j=n^{\frac{4}{5}\alpha }}{\overset{n^{\alpha }}{\sum }}(-1)^{j}%
\binom{n^{\alpha }}{j}(n^{-\frac{\alpha }{2}})^{j}(H(j)-1)| &\leq &\underset{%
j=n^{\frac{4}{5}\alpha }}{\overset{n^{\alpha }}{\sum }}\binom{n^{\alpha }}{j}%
(n^{-\frac{\alpha }{2}})^{j}|H(j)-1| \\
&=&O(n^{\alpha })O(n^{c_{2}})o(e^{-n^{\frac{4}{5}\alpha }})=o(e^{-n^{\frac{%
\alpha }{2}}}).
\end{eqnarray*}

\noindent Thus we get

\[
\Pr (T_{i,u}\text{ is flawed})|_{i\leq 3k\beta \ln n}\leq \Pr (T_{i,u}\text{
is flawed})|_{i=3k\beta \ln n}\approx e^{-n^{\frac{\alpha }{2}}}.
\]

\noindent The remaining part of the proof is identical to that of Lemma 2
for Model RD, and so we are done.

\end{document}